\documentclass[epj]{svjour}
\usepackage{epsfig}
\begin{document}

\title{Primakoff effect in {\boldmath$\eta$}-photoproduction off protons}
\author{A.~Sibirtsev\inst{1,2,3}, J.~Haidenbauer\inst{3,4}, S.~Krewald\inst{3,4} 
and U.-G.~Mei{\ss}ner\inst{1,3,4}
}                     


\institute{
Helmholtz-Institut f\"ur Strahlen- und Kernphysik (Theorie)
und Bethe Center for Theoretical Physics,
Universit\"at Bonn, D-53115 Bonn, Germany  \and
Excited Baryon Analysis Center (EBAC), Thomas Jefferson National
Accelerator
Facility, Newport News, Virginia 23606, USA
\and Institut f\"ur Kernphysik and J\"ulich Center for Hadron Physics,
Forschungszentrum J\"ulich, D-52425 J\"ulich, Germany
\and Institute for Advanced Simulation,
Forschungszentrum J\"ulich, D-52425 J\"ulich, Germany
}

\date{Received: date / Revised version: date}

\abstract{We analyse data on forward $\eta$-meson photoproduction off a 
proton target and extract the $\eta{\to}\gamma\gamma$ decay width 
utilizing the Primakoff effect. The hadronic amplitude that enters into
our analysis is strongly constrained because it is fixed from a global 
fit to available $\gamma{p}{\to}p\eta$ data for differential cross 
sections and polarizations. 
We compare our results with present information on the two-photon $\eta$-decay 
from the literature. We provide predictions for future PrimEx experiments at 
Jefferson Laboratory in order to motivate further studies.
} 

\authorrunning{A. Sibirtsev {\it et al.}}

\maketitle
The decay of the light pseudoscalar mesons into two photons is related to
symmetry breaking through the axial vector anomaly and reveals one of the
fundamental properties of
QCD~\cite{Adler69,Bell69,Bardeen69,Wess71,Witten83,Bijnens93,Goity02,%
Kampf09}. The Adler-Bell-Jackiw (ABJ) 
anomaly~\cite{Adler69,Bell69} allows to determine the decay constant
of those pseudoscalar mesons from the two-photon decay width (invoking a 
smooth extrapolation from the chiral limit to the physical light quark masses). 
While this is indeed feasible for the
$\pi^0$-meson, the situation is more complicated for the $\eta$-meson since the
extrapolation from the $\eta$-mass to zero and the dominance of the 
ABJ anomaly is debatable. Moreover, for the $\eta$ and $\eta^\prime$
mesons 
the decay constants~\cite{Gasser85} alone do not determine the two-photon 
decay widths uniquely. 
It is necessary to know in addition the singlet-octet mixing angle.

In an extension of chiral perturbation theory including the
$\eta^\prime$ meson -- from now on called the
Extended Effective Theory (EET) -- the evaluation of the
$\eta{\to}\gamma\gamma$ decay width requires a proper mixing of the $SU(3)$
pseudoscalar singlet and octet states so that~\cite{Leutwyler98}
\begin{eqnarray}
 \Gamma (\eta{\to}\gamma\gamma) =\frac{\alpha^2 m_\eta^3}{96 \pi^3}
\left[\frac{\cos\theta_P}{f_\eta^8}-\frac{\sqrt{8}\sin \theta_P}{f_\eta^0}
\right]^2 . 
\end{eqnarray}
Here, $f_\eta^0$ and $f_\eta^8$ are the singlet and octet decay constants,
$\theta_P$ is the mixing angle, and $\alpha$ is the fine structure
constant. To estimate the $\eta{\to}\gamma\gamma$
radiative decay width we take from Ref.~\cite{Feldmann00} the following
set of $SU(3)$ parameters: $f_\eta^0$=(1.17$\pm$0.03)$f_{\pi^0}$,
$f_\eta^8$=(1.26$\pm$0.04)$f_{\pi^0}$ and $\theta_P{=}$
${-}(21.2^0{\pm}1.6^0)$. The pion decay constant has the value 
$f_{\pi^0}$=130$\pm$5 MeV and the $\eta$-meson mass is given by the
PDG~\cite{PDG} as $m_\eta{=}547.853{\pm}0.024$~MeV. With these 
parameters the $\eta{\to}\gamma\gamma$ decay width can be estimated to be
\begin{eqnarray}
 0.39 \le \Gamma(\eta{\to}\gamma\gamma) \le 0.52 \,\,\, {\rm keV}.
\label{estimate}
\end{eqnarray}
Alternatively, assuming that $f^0_\eta{=}f^8_\eta{=}f_{\pi^0}$ one can, in
principle, extract the $SU(3)$ mixing angle from the
$\eta{\to}\gamma\gamma$ decay width~\cite{Leutwyler98}.

\begin{figure}[t]
\vspace*{-6mm}
\centerline{\hspace*{5mm}\psfig{file=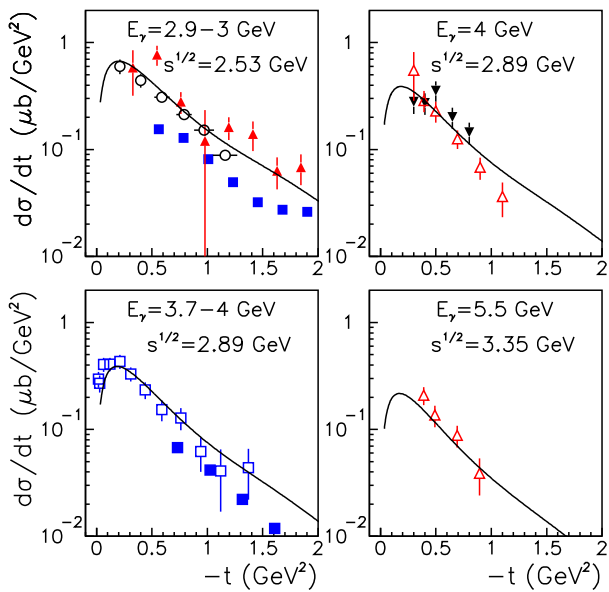,width=9.cm}}
\vspace*{-5mm}
\caption{\label{primet1} Differential cross section 
for the reaction $\gamma{p}{\to}p\eta$ as a function of the 
four-momentum transfer squared at different photon energies 
$E_\gamma$ or invariant collision energies $\sqrt{s}$. 
Data are from
ELSA-Bonn~\cite{Crede05} (filled triangles), 
the Daresbury Electron Synchrotron~\cite{Bussey76} (open circles), 
Cornell~\cite{Dewire71} (inverse triangles),
DESY~\cite{Braunschweig70} (open squares), 
SLAC~\cite{Anderson70} (open triangles), and
CLAS at JLab~\cite{Williams09} (filled squares). 
The lines are the results of our model calculation.}
\end{figure}

Experimentally the $\eta{\to}\gamma\gamma$ decay width was determined
through the QED process $e^+e^-{\to}e^+e^-\eta$ and also from measurements 
of the Primakoff effect with nuclear targets. The results from these two
classes of experiments are in conflict~\cite{PDG}. While all of the QED 
results~\cite{Baru90,Roe90,Williams88,Bartel85,Aihara86,Weinstein83} are in
line with Eq.~(\ref{estimate}) within the experimental error, the Primakoff
measurements~\cite{Browman74,Bemporad67} agree with the EET only within
three times the experimental error bars, say. It was argued~\cite{PDG94} that 
most of the uncertainties in the evaluation from the Primakoff effect are 
due to the nuclear response. 
Recently it was shown~\cite{Rodrigues08} that the $\eta{\to}\gamma\gamma$ 
results might depend significantly on 
contributions from incoherent $\eta$-meson photoproduction off nuclei.

\begin{figure}[t]
\vspace*{-6mm}
\centerline{\hspace*{5mm}\psfig{file=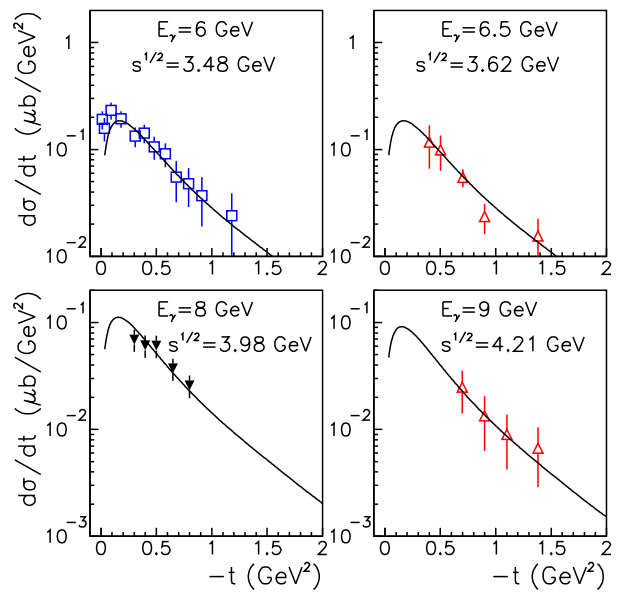,width=9.cm}}
\vspace*{-5mm}
\caption{\label{primb} Differential cross section 
for the reaction $\gamma{p}{\to}p\eta$ as a function of the 
four-momentum transfer squared at different photon energies 
$E_\gamma$ or invariant collision energies $\sqrt{s}$. 
Data are from Cornell~\cite{Dewire71}  (inverse triangles),
 DESY~\cite{Braunschweig70} (open squares) and
SLAC~\cite{Anderson70} (open triangles).
The lines are the results of our model calculation.}
\end{figure}

The Primakoff measurements~\cite{Braunschweig70} on a proton target are
not included in the list of the PDG~\cite{PDG,PDG94}, although the 
quality of these data, collected at DESY, is comparable or even better 
than the results obtained with nuclear targets~\cite{Browman74,Bemporad67}. 
These data were considered~\cite{Laget95} as a strong motivation for further
$\eta{\to}\gamma\gamma$ Primakoff studies~\cite{Primex09} at JLab, 
however, so far they were not used explicitely for an extraction of the 
two-photon decay width.

In the present paper
we utilize the data~\cite{Braunschweig70,Dewire71} available for
forward $\eta$-meson production in the $\gamma{p}{\to}p\eta$ 
reaction at photon energies of 4~GeV and 6~GeV in order to 
extract the $\eta{\to}\gamma\gamma$ decay width. 
To isolate the Primakoff
effect from the hadronic contribution we use the Regge amplitudes
established in our previous study~\cite{Sibirtsev09} of the $\gamma{p}{\to}p\pi^0$
reaction. The free parameters of the helicity amplitudes, namely coupling
constants and form factors, were fixed by a global fit to 
$\eta$-photoproduction data. In the present study we compare our results 
to some $\gamma{p}{\to}p\eta$ data on differential cross 
sections and polarization available at high energies. 
A more thorough comparison to experimental results and predictions at photon 
energies below 3~GeV,
together with more detailed information on the used model parameters,
will be given elsewhere~\cite{Sibirtsev10}.

\begin{figure}[b]
\vspace*{-6mm}
\centerline{\hspace*{5mm}\psfig{file=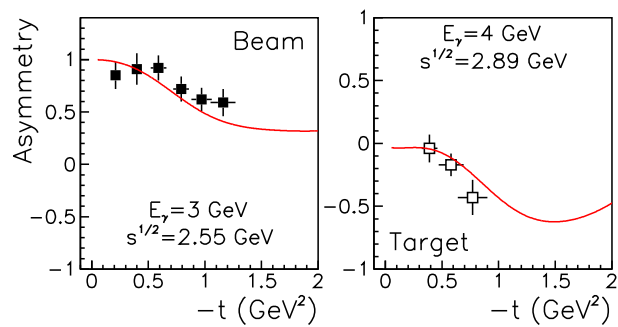,width=9.2cm}}
\vspace*{-42mm}
\caption{\label{primet4} The beam (filled squares) and target (open
squares) asymmetries in photoproduction of $\eta$-mesons from protons
measured~\cite{Bussey76,Bussey81} at the Daresbury Laboratory. 
The lines are the results of our model calculation.}
\end{figure}

Fig.~\ref{primet1} shows $\gamma{p}{\to}p\eta$ differential cross
sections at photon energies from 3~GeV to 5.5~GeV. 
The different measurements available at the same energy are in good 
agreement with each other, considering their uncertainties, 
apart from the recent CLAS results~\cite{Williams09}.
Actually, there are systematic discrepancies between the ELSA-Bonn data
published
in 2005~\cite{Crede05}, in 2008~\cite{Jaegle08} and in 2009~\cite{Crede09},
and the 2009 data from CLAS~\cite{Williams09}. A detailed discussion of 
these discrepancies will be given in Ref.~\cite{Sibirtsev10}.

Differential cross sections for $\eta$-photoproduction 
at energies from 6 to 9~GeV are displayed in Fig.~\ref{primb}. 
The data from the different measurements are described rather well 
by our model calculation.

Data for beam and target asymmetries are presented in 
Fig.~\ref{primet4} as a function of the four-momentum transfer squared. 
Those measurements~\cite{Bussey76,Bussey81} were performed
at the Daresbury Laboratory. It is known~\cite{Bussey76} that
polarization data constitute a rather crucial test for models based on Regge
phenomenology. Indeed, all model calculations \cite{Collins74,Argyres72} 
available at the time when the experiments were published 
failed to reproduce the data on the beam asymmetry. The most 
recent analysis~\cite{Rodrigues08} of the $\gamma{p}{\to}p\eta$ reaction 
is based on a Regge model proposed in 1970~\cite{Braunschweig70}  and
considers only differential cross sections at photon energies above 4~GeV,
but no polarization data. Furthermore, the $\rho$-trajectory
used in Refs.~\cite{Rodrigues08,Braunschweig70} is very different from
the one determined in our global analysis~\cite{Huang09} of the
$\pi^-p{\to}\pi^0n$ 
reaction and from total cross sections for various
other reactions~\cite{Cudell02}. The lines in Fig.~\ref{primet4} are our
results which are clearly in good agreement with the data on beam and 
target asymmetries.

Obviously our Regge model reproduces the $\eta$ photoproduction 
data on differential cross sections and polarizations available at photon 
energies above $\simeq$3~GeV rather well. We take that as confirmation 
for the hadronic part of the reaction amplitude to be reliably determined 
so that it can be used with confidence in the evaluation of the 
Primakoff effect.

The Primakoff effect~\cite{Primakoff51} is due to the one-photon exchange
(OPE) contribution to neutral meson photoproduction. The OPE amplitude
$F_P$ is proportional to the two photon decay width of the $\eta$-meson and
given by
\begin{eqnarray}
F_P = \frac{8 m_p}{t} \sqrt{\frac{\pi
\Gamma(\eta{\to}\gamma\gamma)}{m_\eta^3}}
F_D(t) F_{\eta\gamma\gamma^\ast}(t),
\label{primeff}
\end{eqnarray}
where $m_p$ and $m_\eta$ are the proton and $\eta$-meson masses,
respectively, and $\Gamma$ is the $\eta{\to}\gamma\gamma$ decay width.
$F_D$ and $F_{\eta\gamma\gamma^\ast}$ are form factors at the
$pp\gamma^\ast$ and ${\eta\gamma\gamma^\ast}$ vertices, respectively
($\gamma^\ast$ signifies the virtual (exchanged) photon).
For $F_D(t)$, the Dirac form factor of the proton, we adopt
the parameterization given in Ref.~\cite{Donnachie},
\begin{eqnarray}
F_D (t) = \frac{4m_p^2 - 2.8t}{4m_p^2-t}\frac{1}{(1-t/t_0)^2}
\end{eqnarray}
with $t_0{=}0.71$ GeV$^2$, which is derived under the assumptions
that the Dirac form factor $F_D$ of the neutron and the isoscalar
Pauli form factor vanish and that a dipole form is satisfactory
for $G_M \approx \mu G_E(t)$, cf. \cite{Donnachie}.
There are slight deviations from the dipole form in the
region $-t{<}$0.5 GeV$^2$ we are concerned with here,
cf. for example Ref.~\cite{Belushkin}, but these are
negligible for the present study.
For the form factor $F_{\eta\gamma\gamma^\ast}(t)$ we use
the parameterization given by the CLEO Collaboration in
Ref.~\cite{Gronberg98}:
\begin{eqnarray}
|F_{\eta\gamma\gamma^\ast}(t)|^2=\frac{1}{(4\pi\alpha)^2}
\frac{64\pi\Gamma(\eta{\to}\gamma\gamma)}{m_\eta^3}\frac{1}
{(1-t/\Lambda^2)^2} \ .
\label{etaff}
\end{eqnarray}
A fit to their data~\cite{Gronberg98} yielded the value
$\Lambda{=}$0.774~GeV that is close to the vector dominance model
and to the prediction for the soft nonperturbative region given in
Ref.~\cite{Brodsky81}.
For the Primakoff amplitude as given in Eq.~(\ref{primeff}) we
have to renormalize this form factor so that $F_{\eta\gamma\gamma^\ast}(0){=}1$.
The amplitude of Eq.~(\ref{primeff}) should be
added to the Regge amplitude $F_1$, cf. Refs.~\cite{Sibirtsev09,Sibirtsev10}. 

\begin{figure}[t]
\vspace*{-2mm}
\centerline{\hspace*{3mm}\psfig{file=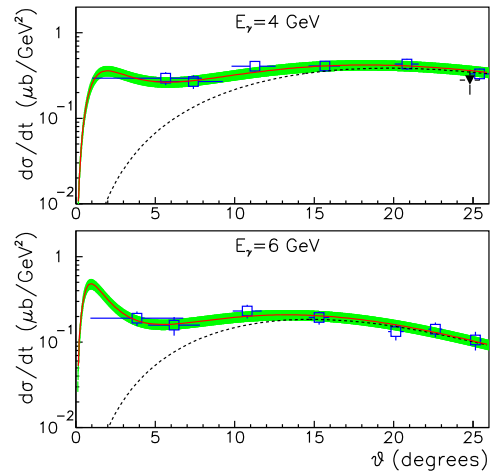,width=7.2cm}}
\vspace*{-4mm}
\caption{Differential cross section 
for the reaction $\gamma{p}{\to}p\eta$ as a 
function of the angle $\vartheta$ in the cm system 
for the photon energies $E_\gamma$=4 and 6~GeV. 
The data are from DESY~\cite{Braunschweig70} (open squares)
and  Cornell~\cite{Dewire71} (inverse triangles). 
The solid (dashed) lines show the results of our Regge calculation 
with (without) inclusion of the OPE contribution.
The shaded band indicates the uncertainty of our results due to the 
hadronic amplitude. 
}
\label{primet1a}
\end{figure}

Since the OPE contributes essentially only at very small $|t|$ it is sensible 
to consider the differential cross section as a function of the center-of-mass
(cm) angle and not of the four-momentum transfer squared. This is
done in Fig.~\ref{primet1a} where we include data from DESY at energies 
of 4~GeV and 6~GeV~\cite{Braunschweig70} (open squares) and
from Cornell at 4~GeV~\cite{Dewire71} (inverse triangles).
The dashed lines indicate our calculations without the OPE amplitude.
It is clear that at angles below 10$^\circ$ the data are underestimated,
which is a direct indication for the required additional contribution 
due to the Primakoff amplitude, Eq.~(\ref{primeff}). 

Let us now fit the data, taking the $\eta{\to}\gamma\gamma$ radiative decay
width as free parameter. Unfortunately, there is no information with regard 
to the angular resolution of the
measurements~\cite{Braunschweig70,Dewire71}. Only the intervals
of the four-momentum squared where the DESY results were obtained are
known and those are indicated in Fig.~\ref{primet1a} as an uncertainty in
the cm angle $\vartheta$. Fortunately, below $\vartheta{\le}15^\circ$ the 
cross-section data exhibit practically no angular dependence so that these 
uncertainties have no significant influence on our solution. We 
obtain for the decay width the values 
$\Gamma(\eta{\to}\gamma\gamma)$=0.86$\pm$0.11~keV at the photon energy 
of 4~GeV and $\Gamma(\eta{\to}\gamma\gamma)$=0.70$\pm$0.09~keV 
at $E_\gamma$=6~GeV. The fit includes all
data points shown in Fig.~\ref{primet1a}. A somewhat unpleasant finding 
is that for both photon energies the fit results in a rather low $\chi^2$,
namely $\chi^2/{\rm data~ point}{\simeq}$0.3. For a statistically uncorrelated 
set of data points one would expect a value around 1. 
The shaded band in Fig.~\ref{primet1a} indicates the uncertainty in the
employed hadronic amplitude \cite{Sibirtsev10}. 

\begin{figure}[t]
\vspace*{-3mm}
\centerline{\hspace*{3mm}\psfig{file=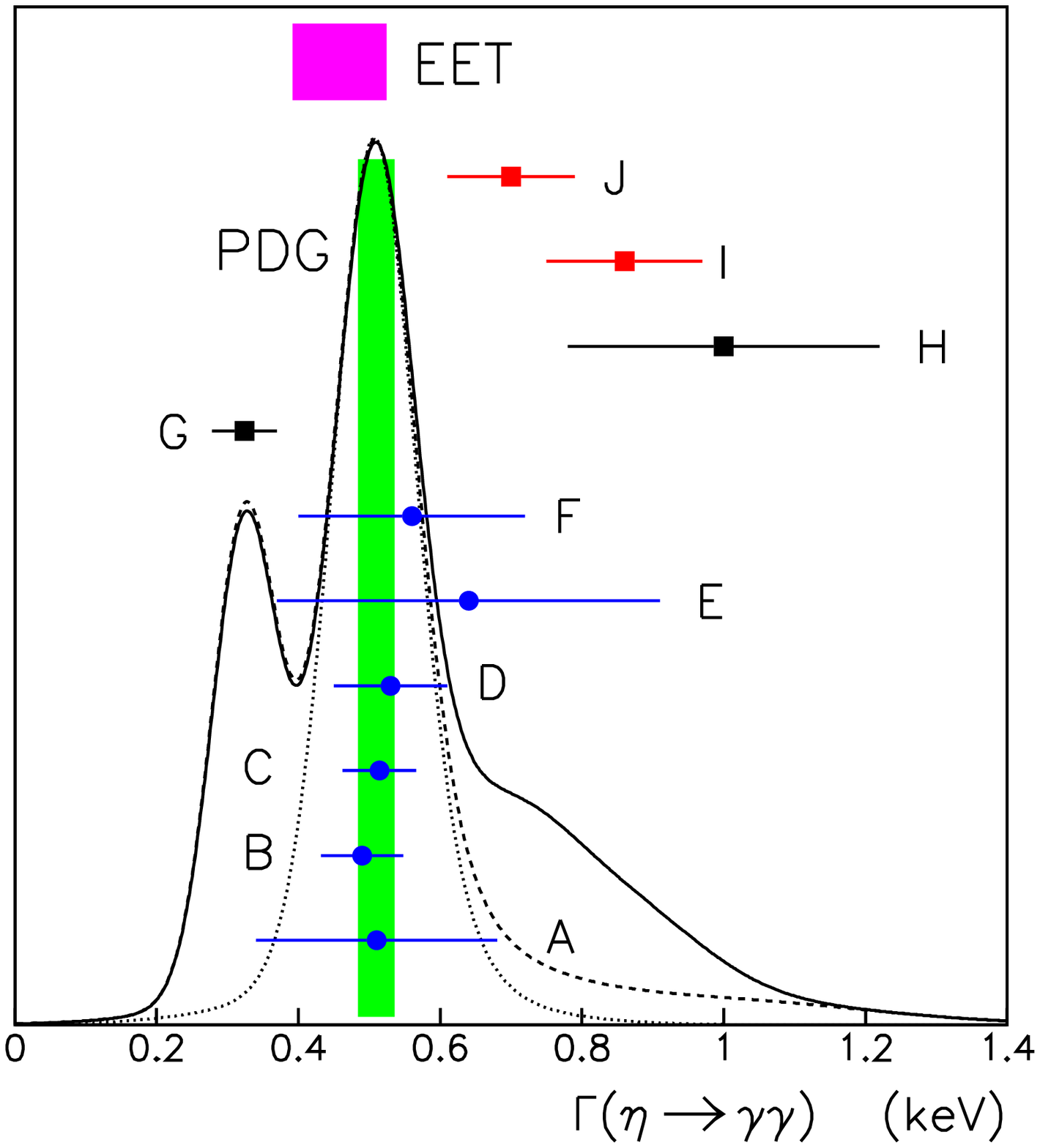,width=7.2cm}}
\vspace*{-3mm}
\caption{The $\eta{\to}\gamma\gamma$ radiative decay width from different
analyses. The results are taken from: A~\cite{Baru90}, B~\cite{Roe90},
C~\cite{Williams88}, D~\cite{Bartel85}, E~\cite{Aihara86},
F~\cite{Weinstein83}, G~\cite{Browman74}, H~\cite{Bemporad67}. 
Our own results obtained from a fit of the $\gamma{p}{\to}\eta{p}$
differential cross sections~\cite{Dewire71,Braunschweig70} at the photon
energies of 4 and 6 GeV, are denoted by I and J, respectively. 
Circles are data obtained from the $e^+e^-{\to}e^+e^-\eta$ reaction, while 
squares indicate results obtained by Primakoff-effect measurements. 
The shaded box, indicated as PDG, is an averaged result~\cite{PDG}. 
The box indicated as EET shows the limit given by Eq.~(\ref{estimate}). 
The lines indicate data distribution functions as explained in the text.
}
\label{primet9}
\end{figure}

Keeping in mind possible ambiguities of our results due to the unknown 
resolution we now compare the $\eta{\to}\gamma\gamma$ 
radiative decay width evaluated
here with the world data. Fig.~\ref{primet9} contains results from
different measurements available in the literature. Specifically, circles 
indicate data~\cite{Baru90,Roe90,Williams88,Bartel85,Aihara86,Weinstein83} 
determined from the $e^+e^-{\to}e^+e^-\eta$ reaction, while the squares 
are results~\cite{Browman74,Bemporad67} obtained with the
Primakoff effect from measurements on nuclear targets. A detailed 
discussion of the various measurements is given in Ref.~\cite{PDG94}. 

Recently the results from the Cornell measurement~\cite{Browman74},
indicated by G in Fig.~\ref{primet9}, were 
re-analyzed~\cite{Rodrigues08} assuming different contributions from
incoherent $\eta$-meson photoproduction from nuclei which led to an
$\eta{\to}\gamma\gamma$ radiative decay width of 0.476~keV. This
illustrated that the analysis of the data~\cite{Browman74} 
depends significantly on the incoherent background at angles below 
$\simeq$2$^0$ in the laboratory frame and on the model applied. 
Clearly the data obtained~\cite{Dewire71,Braunschweig70} with a 
proton target are clean with respect to such background contributions 
to the Primakoff effect so that our evaluation here should be more reliable. 
Therefore, we agree with the conclusion of Laget~\cite{Laget95} that 
future measurements with a proton target could be quite promising.

Our results obtained for photon energies of 4 and 6~GeV are indicated
in Fig.~\ref{primet9} as I and J, respectively\footnote{We remark that the
  effects of the form factors in Eq.~(\ref{primeff}) is small, setting e.g. 
  $F_{\eta\gamma\gamma^\ast}(t)$
  to one leads to a few percent shift in the central value for the width,
  well within the uncertainty induced from the hadronic amplitudes.}. 
We see that there is a
partial conflict between these results and data obtained from the 
$e^+e^-{\to}e^+e^-\eta$ measurements. The shaded band in 
Fig.~\ref{primet9} illustrates the averaged result from the PDG~\cite{PDG},
based on the measurements A to D. The lines show 
data distribution functions that were obtained in the following
way~\cite{PDG}: To each measurement shown in Fig.~\ref{primet9} a Gaussian 
distribution is assigned with a central value, 
and a dispersion given by the error bar and the integral area proportional to 
the inverse error bar. The dotted line in Fig.~\ref{primet9} represents the 
sum of the Gaussians for the measurements A to D. 
The dashed line indicates the sum of the measurements A to H. The solid 
line is the sum obtained with all data, including our results I and J.  

From this data distribution analysis we conclude that even when taking 
into account all measurements and considering the correction
proposed~\cite{Rodrigues08} for G, it is still difficult to infer 
that the $\eta{\to}\gamma\gamma$ radiative decay width is known now 
with high accuracy, say 5\%, as given by the PDG~\cite{PDG}. 
Note that the recent averaged value of $\Gamma(\eta{\to}\gamma\gamma)$ 
of the PDG was obtained by neglecting the $e^+e^-{\to}e^+e^-\eta$
measurements~\cite{Aihara86,Weinstein83} indicated as E and F in
Fig.~\ref{primet9}. In addition, the results G and H obtained by the
Primakoff measurement were always neglected by the PDG~\cite{PDG} as 
recommended in Ref.~\cite{Nefkens02} without any solid argumentation,
solely on the basis that these results are in disagreement with 
other measurements.

\begin{figure}[t]
\vspace*{-6mm}
\centerline{\hspace*{3mm}\psfig{file=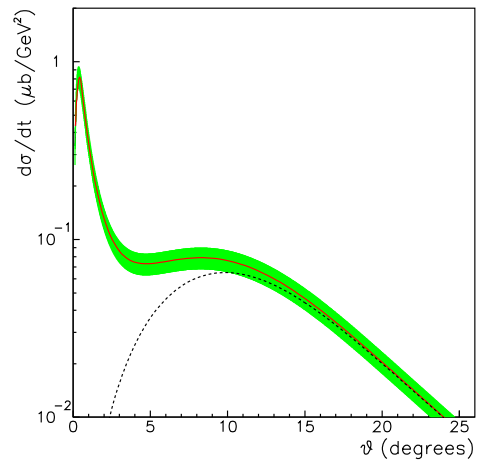,width=7.cm}}
\vspace*{-4mm}
\caption{Differential cross section 
for the reaction $\gamma{p}{\to}p\eta$ as a 
function of the angle $\vartheta$ in the cm system at the photon energy
$E_\gamma$=11 GeV. The solid (dashed) lines show the results of our Regge
calculations with (without) inclusion of the one-photon-exchange contribution.
The shaded band indicates the uncertainty of our prediction due to the 
hadronic amplitude. 
}
\label{primet2a}
\end{figure}

In Fig.~\ref{primet2a} we display the differential cross section at forward 
angles for $\eta$-meson photoproduction at the photon energy 11~GeV. This
energy was suggested by the PrimEx Collaboration for a future
experiment in Hall D at JLab~\cite{Primex09}. 
Here the dashed line is the hadronic contribution
alone, while the solid line accounts for the sum of the OPE and the hadronic
amplitude. The shaded band indicates the uncertainty of our prediction due to
the hadronic amplitude. Note that for this prediction we used the standard
value $\Gamma(\eta{\to}\gamma\gamma)$=0.51 keV \cite{PDG}.
Obviously also at higher energies the signal from the 
Primakoff effect is much larger than the uncertainty in the hadronic amplitude,
even up to angles around $\vartheta\approx 5^\circ$,
Thus, we believe that it is rather promising to perform further high
precision measurements of the two-photon decay of the $\eta$-meson utilizing
the Primakoff effect as proposed by the PrimEx Collaboration at
JLab~\cite{Primex09}.

 \begin{acknowledgement}
This work is partially supported by the Helmholtz Association through funds
provided to the virtual institute ``Spin and strong QCD'' (VH-VI-231), by
the EU Integrated Infrastructure Initiative  HadronPhysics2 Project (WP4
QCDnet) and by DFG (SFB/TR 16, ``Subnuclear Structure of Matter''). This
work was also supported in part by U.S. DOE Contract No. DE-AC05-06OR23177,
under which Jefferson Science Associates, LLC, operates Jefferson Lab. A.S.
acknowledges support by the JLab grant SURA-06-C0452  and the COSY FFE
grant No. 41760632 (COSY-085). 
\end{acknowledgement}


\end{document}